\documentclass{aastex63}
\usepackage{graphicx} 
\newcommand{\msun}{${\rm M_{\odot}}$ }
\newcommand{\bvf}{Brunt-V\"ais\"al\"a }

\begin{document}

\title{A systematic study of the connection between white dwarf period spectra and model structure}
\author{Agn\`es Bischoff-Kim }
\affiliation{Penn State Scranton \\
120 Ridgeview Drive \\
Dunmore, PA 18512, USA}

\begin{abstract}
    To date, pulsational variability has been measured from nearly 70 DBVs and 500 DAVs, with only a fraction of these having been the subjects of asteroseismic analysis. One way to approach white dwarf asteroseismology is forward modeling, where one assumes an interior structure and calculates the model's periods. Many such models are calculated, in the search for the one that best matches an observed period spectrum. It is not computationaly manageable, nor necessary, to vary every possible parameter for every object. We engage in a systematic study, based on a sample of 14 hydrogen atmosphere white dwarfs, chosen to be representative of the types of pulsation spectra we encounter in white dwarf asteroseismology. These white dwarfs are modeled with carbon and oxygen cores . Our goal is to draw a connection between the period spectra and what parameters they are most sensitive to. We find that the presence of longer period modes generally muddies the mass and effective temperature determinations, unless continuous sequences of $\ell=1$ and $\ell=2$ modes are present. All period spectra are sensitive to structure in the helium and hydrogen envelope and most to at least some fatures of the oxygen abundance profile. Such sensitivity can be achieved either by the presence of specific low radial overtone modes, or by the presence of longer period modes. Convective efficiency only matters when fitting periods greater than ~800 s. The results of this study can be used to inform parameter selection and pave the way to pipeline asteroseismic fitting of white dwarfs.
\end{abstract}

\section{Astrophysical Context}
\label{sec:introduction}

White dwarfs are the end result of the evolution of stars with masses less than 10 or 11 ${\rm M_{\odot}}$ \citep[e.g.][]{Woosley15} and as such, hold in their interiors the fossil records of the evolution of the vast majority of the stars in our galaxy ($\sim$ 98\%). The resulting chemical profiles help constrain physical processes such as nuclear fusion, core overshooting, mass loss and diffusion. Most white dwarfs fall in two spectral classes, DA and DB, with hydrogen and helium dominated atmospheres respectively. Pulsating white dwarfs have a "V" designation (e.g. DAV or DBV). In this work we focus on DAV's, even though most of the results also apply to DBV's. We also consider exclusively carbon-oxygen core white dwarfs. This is the internal composition expected for white dwarfs ranging in mass between $\sim 0.45$ and $1 \; {\rm M_\odot}$. Above that mass, we expect oxygen-neon cores, and below $0.45 {M_\odot}$, we expect helium cores \citep[e.g.][]{Corsico19}. This mass range comprises the vast majority of white dwarfs, with the mass distribution of DAs strongly peaking at an average mass of $\sim 0.65 \; {\rm M_\odot}$ \citep{Kepler17}.

Pulsations observed in DAs and DBs are driven in the convection zone and are $g$-modes. Because of geometric cancellations, we do not expect to observe modes past $\ell$=2 \citep{Dziembowski77}. Modes are also described by their radial overtone. In this paper, we shall use $k$ to denote that number. Another notation often used in the literature is $n$. Most observed white dwarf pulsation spectra consist of fewer than a dozen modes. Asteroseismic fitting of these spectra allows us to infer interior structure of white dwarfs. This in turn helps constrain the physical processes involved in stellar evolution calculations. While there has been a wealth of studies of pulsating white dwarfs with tantalizing results \citep[e.g.][]{Metcalfe03b,Hermes17,Giammichele18}, we yet lack a coherent picture of their internal structure determined from fitting the pulsation spectra.

To date, we have obtained pulsation spectra for 500 hydrogen atmosphere white dwarfs (DAV's) and nearly 70 helium atmosphere white dwarfs \citep{Romero22,Vanderbosch22}. Only a fraction of these have been the subject of asteroseismic fitting because we have not developed pipeline fitting for these objects yet. The largest efforts to date include that of \citet{Castanheira09}, with the study of 83 DAV's. The authors assumed cores made up of a homogeneous, 50/50 oxygen-carbon mix. From stellar evolution calculations, we know that this is not physical \citep[e.g.][]{Corsico19}. \citet{Hall23} performed the asteroseismic fitting of 29 DAVs observed with the Kepler and K2 missions. More physical, but still fixed, oxygen profiles were considered. The authors showed that for at least one of the stars in the study, the period spectrum was sensitive to core structure. We revisit some of these objects in this work. Both studies were performed using the White Dwarf Evolution Code (WDEC), where fixed core compositions go in as input (forward modeling). Another major effort was the study of 74 newly discovered DAVs in TESS data \citep{Romero22}, and ten years earlier, 44 DAVs by the same group \citep{Romero12}. The latter studies are based on fully evolutionary models. While some aspects of envelope structure, for instance the thickness of the hydrogen layer can be varied by setting the number of thermal pulses on the AGB, much of the internal chemical profiles are determined by how the relevant physical processes are implemented in the models. In the present paper, we use the WDEC and focus on the parameterization used in that code. But some results, in particular those that pertain to mass, effective temperature, and the thickness of the hydrogen layers may be of use to studies that use fully evolutionary models. 

For forward modeling in particular, it is useful to know which parts of the model the observed pulsations spectra are sensitive to, as it is computationally onerous to try and vary every possible parameter (the WDEC in its current version includes 15 parameters). From the theory of non-radial oscillations we expect lower radial overtone modes (lower $k$) to have their resonant cavity deeper in the star and so would expect them to be more sensitive to core structure, while higher $k$ modes should respond more strongly to the envelope \citep{Unno89}. This, however has been shown independently to not be a solid rule, for white dwarfs, as well as for sub-dwarfs \citep{Bischoff-Kim17a,Charpinet17a}. As a result, we cannot simply rely on this rule of thumb. And yet, we do expect a connection between the periods present in the spectrum of a white dwarf and its sensitivity to different parameters. We explore such connections in a systematic study and work toward a goal of mapping periods spectrum types to parameter sensitivity.

We motivate our selection of objects for the study and describe the asteroseismic modeling and fitting in section \ref{sec:methods}. We present our results in section \ref{sec:results} then discuss them in section \ref{sec:discussion}. We highlight important points and results in section \ref{sec:conclusions}.

\section{Method}
\label{sec:methods}
\subsection{Object selection}
\label{sec:object_selection}
For this numerical experiment, we selected 14 DAV's. We chose a number of them from \citet{Hermes17}, and others from older, ground based observing campaigns. We used the published white dwarf tables of \citet{Bognar16} and references therein for the legacy objects. When there was more than one paper reporting on the pulsations of an object, we collated all the periods found. If the same mode was observed in more than one observing campaign, as was often the case, we picked the most recent period reported. We should emphasize, however, that \emph{the periods themselves were not used in the fits}, but rather were used as a basis to pick a set of periods from a fiducial model. This means that the exact value of the period is not relevant for this study. What is more important is its $\ell$ and $k$ identification. 

We list the objects, along with a broad description of their period spectra, in table \ref{tab:objects}. "Short periods" are defined as periods under 500~s ($k<8$ for our fiducial model), while longer period modes obey 500~s and above. Furthermore, an asterisk in that column indicates the presence of modes over 800~s. We grouped objects with similar pulsation spectra under different types, defined for the purpose of this work. For all types, we have more than 1 object. 

    \begin{table}
  \begin{center}
  \caption{List of test objects. Long periods are defined as $\rm{P} \gtrsim 500$ s. An asterisk indicates the presence of periods longer than 800~s. For "EPIC" objects, mass and effective temperatures are from \citet{Hermes17}. For the other stars, we used the spectroscopy by \citet{Gianninas11}.
     \label{tab:objects}
}
  \begin{tabular}{ccccccccc}
  \hline 
    Object & ${\rm T_{eff}}$ & Mass & Number& $\ell=1$& $\ell=2$& Short   & Long & Type  \\
           &                 &      & of modes & modes & modes &  periods  &  periods &  \\
    \hline
    EPIC 7595781    & 12650~K  & 0.67 ${\rm M_\odot}$&   12  &   Yes &   Yes &   Yes &   Yes* &   Type 1  \\
    EPIC 201719578  & 11070~K  & 0.57 ${\rm M_\odot}$&   23  &       &       &       &        &           \\
    \hline
    GD 66           & 12430~K  & 0.68 ${\rm M_\odot}$&   7   &   Yes &   Yes &   Yes &   Yes  &   Type 4  \\
    G 185-32        & 12600~K  & 0.68 ${\rm M_\odot}$&  8   &       &       &       &        &           \\
    \hline
    EPIC 211914185  & 13590~K  & 0.88 ${\rm M_\odot}$&  2   &   Yes &   No  &   Yes &   No   &   Type 2  \\
    G 117-B15A      & 12240~K  & 0.69 ${\rm M_\odot}$&  3   &       &       &       &        &           \\
    \hline
    EPIC 220258806  & 12800~K  & 0.66 ${\rm M_\odot}$&  11  &   Yes &   Yes &   Yes &   No   &   Type 3  \\
    R 548           & 12480~K  & 0.64 ${\rm M_\odot}$&  6   &       &       &       &        &           \\
    L19-2           & 12330~K  & 0.71 ${\rm M_\odot}$&  5   &       &       &       &        &           \\
    GD 165          & 12440~K  & 0.69 ${\rm M_\odot}$&  6   &       &       &       &        &           \\
    \hline
    EPIC 220204626  & 11620~K  & 0.71 ${\rm M_\odot}$& 7   &   Yes &   Yes &   No  &   Yes  &   Type 5  \\    
    EPIC 206212611  & 10830~K  & 0.60 ${\rm M_\odot}$& 3   &       &       &       &        &           \\
    \hline
    EPIC 210397465  & 11200~K  & 0.45 ${\rm M_\odot}$& 11  &   Yes &   Yes &   No  &   Yes* &   Type 6a \\
    EPIC 229228364  & 11030~K  & 0.62 ${\rm M_\odot}$& 3   &   Yes &   No  &   No  &   Yes* &   Type 6b \\
    \hline
 \end{tabular}
 \end{center}
\vspace{1mm}
\end{table}

\subsection{Mode selection}
For the periods present in the spectra of each object, we need to choose an $\ell$ identification. $\ell$ identification may come from the presence of multiplets. When that was the case, the authors cited in section \ref{sec:object_selection} offered their $\ell$ identifications. We adopted those values whenever available. In the tables in the appendix, such modes are identified with an asterisk. Most modes, however, are not observed to be members of multiplets and so we have to make assumptions for the reminder of them. As higher $\ell$ modes suffer from geometric cancellations \citep{Dziembowski77}, the usual assumption is that modes that have amplitudes large enough to be observed are $\ell=1$ modes. One common and reasonable starting point is to assign $\ell=1$ to the higher amplitude modes, then look at the remaining modes. Any mode too close to a high amplitude mode for any reasonable period spacing matching the expected mass and effective temperature of the white dwarf has to be $\ell=2$. The method can then be refined by trying every reasonable combinations. In the present work, we did not refine the method, as the goal was not to obtain a best fit to the objects, but rather use representative sets of periods from the fiducial model to perform numerical experiments. We also note that what has a greater impact on how sensitive the period spectra are to any given parameter is the period of the mode (long versus short) and not its $\ell$ identification. 

\subsection{Asteroseismic fitting}
\label{sec:fitting}

\begin{figure}[h!]
\epsscale{0.6}
\plotone{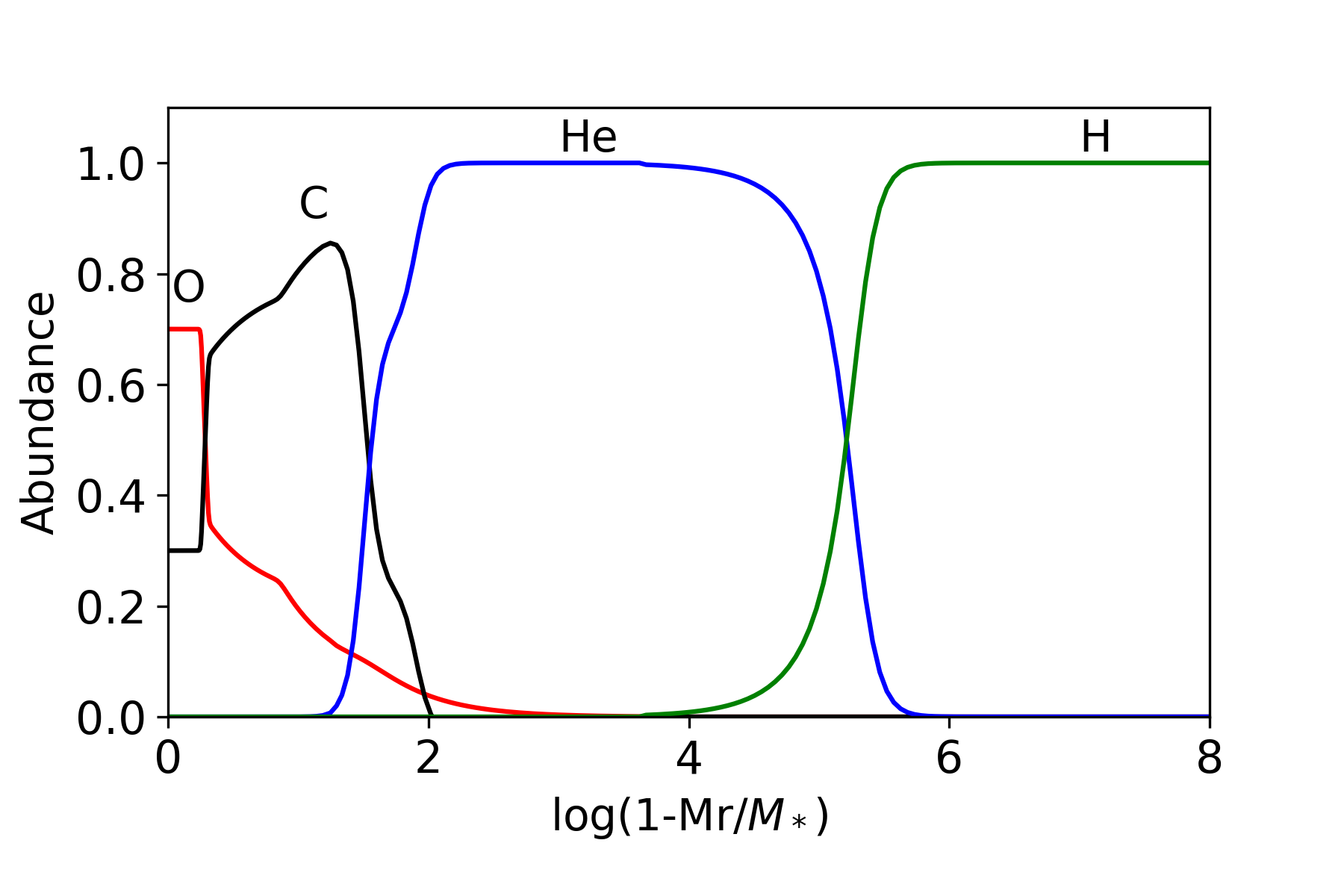}
\caption{Chemical composition profiles for the fiducial model. The surface of the model is on the right.  
\label{fig:comprofiles}
}
\end{figure}

We calculated the models required for this work using the White Dwarf Evolution Code, WDEC \citep{Bischoff-Kim18a}. The code is open source and may be obtained from GitHub (Codebase: https://github.com/kim554/wdec). A key feature of WDEC is the ability to vary the interior chemical profiles. Instead of using time dependent diffusion to calculate core chemical profiles based on some starting chemical composition, the code instead accepts the profiles as an input and then calculates a model that satisfies the equations of stellar structure and calculates the associated non-radial oscillation modes. The code allows to vary a maximum of 15 parameters, described in detail in \citet{Bischoff-Kim08c} and \citet{Kim20}. We refer the reader to those two publications for the details and summarize the key features of those 15 parameters in table \ref{tab:parameters}. For reference, we add a column that shows the range of values that are physicaly reasonable to use when computing models with the WDEC. The values are informed by stellar evolution calculations. While we parameterized the efficiency of convection for this work, WDEC can set this parameter based on 3D convection models \citep{Tremblay15} and non-linear lightcurve fitting \citep{Provencal15}. \citet{Bischoff-Kim08a} discuss at length how MLT convection is treated in WDEC. An example interior chemical composition profile is shown in Fig. \ref{fig:comprofiles}. 

\begin{table}
  \begin{center}
  \caption{The 15 parameters that can be varied in the White Dwarf Evolution Code
     \label{tab:parameters}
}
  \begin{tabular}{llll}
  \hline 
    Parameter & Short description & Fiducial & Reasonable \\
              &                   & value    & range  \\
    \hline
    \multicolumn{2}{l}{\textbf{Bulk properties of the model}}                   &               &               \\
    ${\rm T_{eff}}$ & Effective temperature [K]                                 & 11000         & 10000 - 14000 \\       
    Mass & Stellar mass (solar)                                                 & 0.6           & 0.35 - 1.00   \\
    \hline
    \multicolumn{2}{l}{\textbf{He and H envelope parameters}}                   &               &               \\
    ${\rm M_{env}}$ & Location of the base of the helium layer (-log)           & 1.6           & 1.2 - 3.0     \\
    ${\rm M_{He}}$ & Location of the base of the pure helium layer (-log)       & 2.0           & 2.0 - 4.0     \\
    ${\rm M_{H}}$ & Location of the base of the hydrogen layer (-log)           & 5.0           & 3.6 - 7.0     \\
    Xhebar  &   Helium abundance in the homogeneous He/H region                 & 0.7           & 0.5 - 9.0     \\
    alph1   & Sharpness of the transition at the base of the helium layer       & 16            & 4 - 20        \\
    alph2   & Sharpness of the transition to pure helium                        & 16            & 4 - 20        \\
    alpha   & Efficiency parameter for MLT convection \citep{Bohm-Vitense58}    & 0.65          & N/A           \\
    \hline
    \multicolumn{2}{l}{\textbf{C/O core parameters}}                            &               &               \\
    h1  & Central oxygen abundance                                              & 0.7           & 0.6 - 0.8     \\
    h2 & Oxygen abundance in the "lower plateau" (fraction of h1)               & 0.5           & 0.5 - 0.9     \\ 
    h3 & Oxygen abundance near the edge of the C/O core (fraction of h2)        & 0.7           & 0.7 - 0.85    \\
    w1 & Location of the edge of the homogeneous C/O region (${\rm M_r/M_*}$)   & 0.45          & 0.35 - 0.45   \\
    w2 & Location of the second "knee" (${\rm M_r/M_*}$)                        & 0.12          & 0.10 - 0.65   \\
    w3 & Location of the edge of the C/O core (${\rm M_r/M_*}$)                 & 0.35          & 0.2 - 0.40    \\
    \hline
 \end{tabular}
 \end{center}
\vspace{1mm}
\end{table}

We picked a fiducial model in the middle of the ZZ Ceti instability strip and with canonical chemical profiles (Fig. \ref{fig:comprofiles}). The parameters of the fiducial model are listed in table \ref{tab:parameters}. We then ran 15 series of models, varying one parameter at a time over the full allowed range of that parameter, in steps fine enough to give us a clear picture of how the goodness of fit varied as we scanned all possible or reasonable values of the parameter. 

To perform the numerical experiment, we calculated the periods of the fiducial model, and selected subsets of periods from that list, to best match the observed period spectrums of the DAV's listed in table \ref{tab:parameters}. The periods of the fiducial model, along with the selected subset of period for each star is listed in the appendix. We performed a period match and a computation of the quality of fit parameter according to the method we use when performing the asteroseismic fit of an actual star. The goodness of fit parameter is computed using the formula

\begin{eqnarray}
\label{eq:fiteq1}
\sigma_{\rm RMS} = \sqrt{\frac{1}{W} \sum_{1}^{n_{\rm obs}} {w_i(P^{\rm calc}_i-P^{\rm obs}_i)^2}}, \\
W=\frac{n_{\rm obs}-1}{n_{\rm obs}}\sum_{1}^{n_{\rm obs}}w_i
\end{eqnarray}

\noindent where $n_{\rm obs}$ is the number of periods present in the pulsation spectrum and the weights $w_i$ are the inverse square of the errors on mode $i$. Since we are not working with observed pulsation spectra, we have to choose the weights. \citet{Hermes17} noted a clear dichotomy between modes shorter than 800~s and those longer. In their sample of 27 DAVs, they found that the longer period modes had an error of $\sim 2 \mu {\rm Hz}$  on the average. On the flip side, shorter period modes, can be coherent down to a small (and variable) fraction of a $\mu {\rm Hz}$. In order to reflect that dichotomy while not neglecting the longer period modes, we weighed the shorter period modes 10 times more than the longer period modes.

\section{Results}
\label{sec:results}
\begin{figure}[h!]
\epsscale{1.}
\plotone{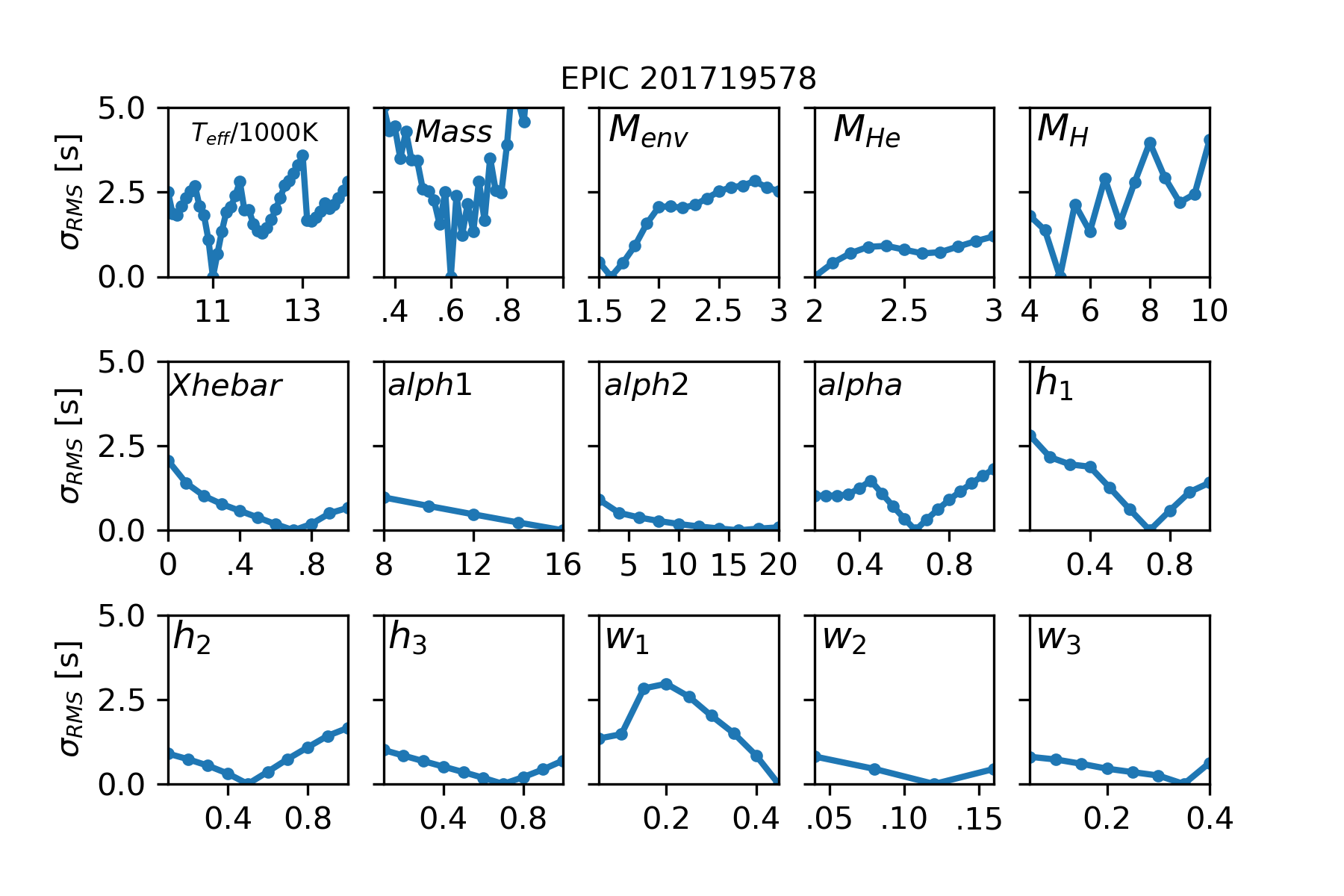}
\caption{Variation of the quality of fit as a function of the value of each of 15 parameters considered in this work. Such a diagram was produced for each object in table \ref{tab:objects}. The one shown here (for EPIC 201719578) is representative. 
\label{fig:pardep}
}
\end{figure}

We performed the calculations described in section \ref{sec:fitting} for each object listed in table \ref{tab:objects} and for each, produced graphs like those shown in Fig. \ref{fig:pardep}. With 14 such figures, we found patterns detailed in this section.

\subsection{Effective temperature}

The dependence of the quality of fit on varying effective temperatures falls into two categories: a) there is a unique, clearly defined absolute minimum or b) there are multiple minima, with one absolute minimum (a sawtooth pattern dipping to a unique solution).  Types 2 and 3 (see table \ref{tab:objects}) exhibit pattern a) while the others exhibit pattern b). It appears that the presence of long period modes muddies the temperature determination by giving rise to multiple minima. 

\subsection{Stellar mass}

A similar pattern occurs for stellar mass, except there are cases when there is no clearly defined, unique absolute minimum. The sawtooth pattern is nearly flat (Fig. \ref{fig:periodrifts_mass} and \ref{fig:thetwotype1s_mass}). The only period spectra that lead to a clear, unique determination of stellar mass are type 3. Types 4 and 6, as well as EPIC 7595781 (Type 1) have the least well determined mass. It appears that for mass determination, it is beneficial to have $\ell=2$ modes, and that again, longer period modes muddy the waters.

\subsection{Location of transitions in the Helium envelope}

The envelope mass ${\rm M_{env}}$ was clearly determined for type 2 pulsation spectra and to a lesser extent, type 3. There was still a clear absolute minimum for the other types as well. This is a parameter that is universally well determined. The same is true of the other helium envelope parameter, ${\rm M_{He}}$.

\subsection{Other Helium envelope parameters}

Types 1, 5, and 6 are more sensitive to the parameter Xhebar. All period spectra considered are sensitive to this parameter even if to a lesser extent. Types 2 and 3 are completely insensitive to the parameter alph1. The other pulsation spectra are sensitive to that parameter, but less so than to xhebar. We find that none of the period spectra are sensitive to alph2.

\subsection{Hydrogen envelope}

For ${\rm M_{H}}$ types 2 and 3 have the most defined absolute minima. Of note, EPIC 211914185's spectrum is very sensitive to the hydrogen layer mass. EPIC 211914185 only has two modes, $k=1$ and $k=2$. All types are sensitive to this parameter, even if types 4, 5 and 6 are less well defined. One thing we observe for most (but not all) period spectra considered is that the fits are insensitive to any hydrogen layer thinner than $\sim 10^{-7}$.

\subsection{Convective efficiency}

Types 2 and 3 are completely unresponsive to the value of the parameter that dictates convective efficiency (alpha). Types 1 and 5 are sensitive at a level comparable to other parameters. EPIC 220204626 is completely insensitive to the convective efficiency. Unlike the type 1 and 5 stars, its longest period modes are below 800~s.

\subsection{Core parameters}

All period spectra are sensitive to some degree to h1 and h2. With the striking exception of EPIC 211914185, all also respond to w1. The level of sensitivity is similar to the response to ${\rm M_{env}}$ or Xhebar. When it comes to core parameters, G 117-B15A and EPIC 211914185 look completely different. As mentioned before, EPIC 211914185's two modes are $k=1$ and $k=2$, while G 117-B15A has three modes, $k=2$, 3 and 4. The former is mostly insensitive to core structure, while G 117-B15A's spectrum responds strongly to core structure. The other core parameter (h3, w2, and w3) elicit a very weak response, if any from all period spectra considered. The G 117-B15A-like period spectrum has the strongest response to the core parameters of all. 

\section{Discussion}
\label{sec:discussion}

\subsection{Mass and effective temperature}
\label{sec:summary_massteff}
\begin{figure}[h!]
\epsscale{0.6}
\plotone{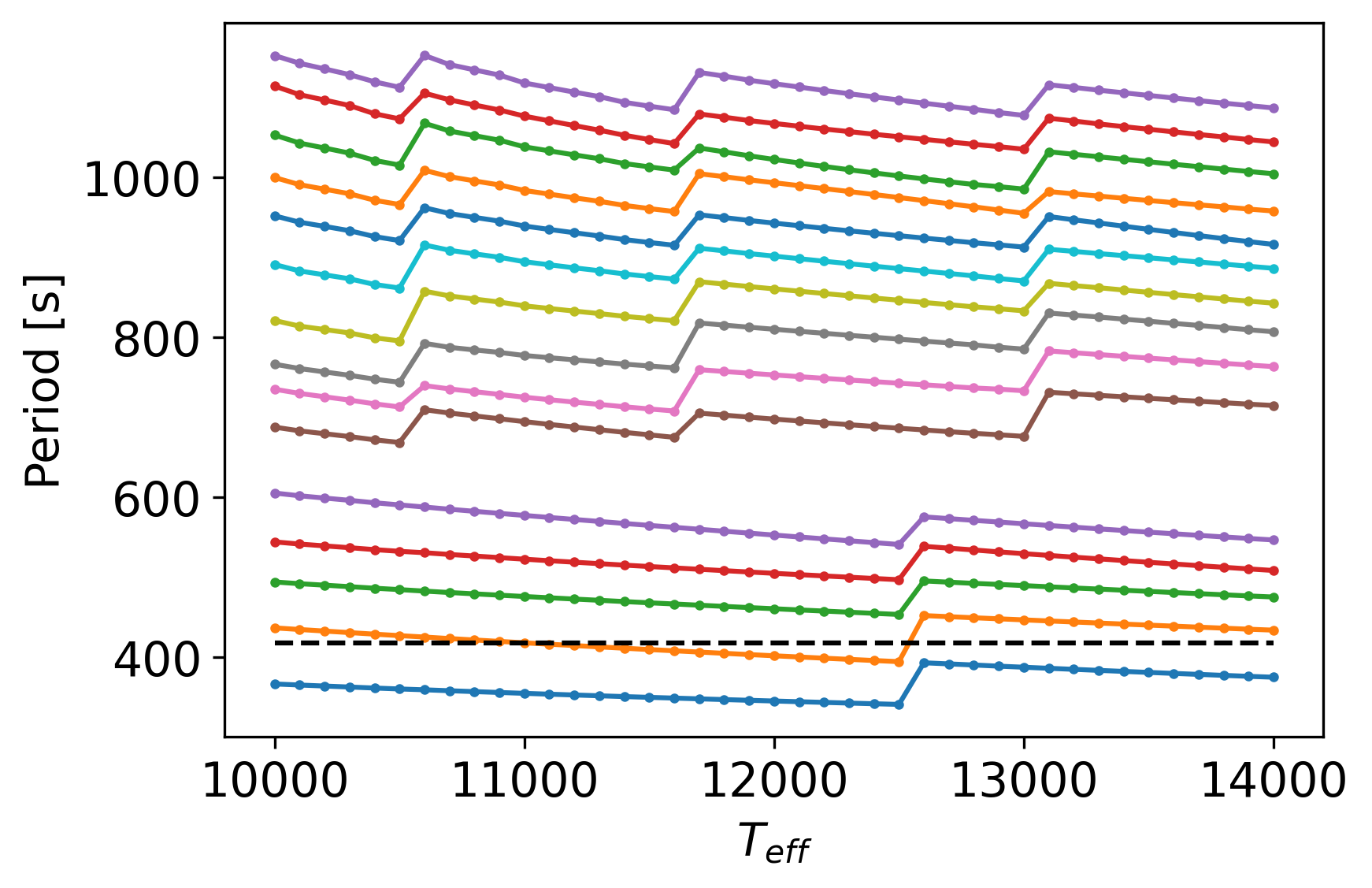}
\caption{Evolution of the best fit period for each mode of EPIC 201719578 as a function of effective temperature. We chose the object because of its rich period spectrum (23 periods). The horizontal dashed line around 400~s marks the location of one of the modes present in the star. The period fitting algorithm select periods to stay close to that line. It may look like the jump around 12600~K happens too soon, but the algorithm has to optimize such match for all the periods, not just one. \textbf{For clarity, only the $\ell=1$ modes are shown.}
\label{fig:periodrifts}
}
\end{figure}

\begin{figure}[h!]
\epsscale{1.}
\plotone{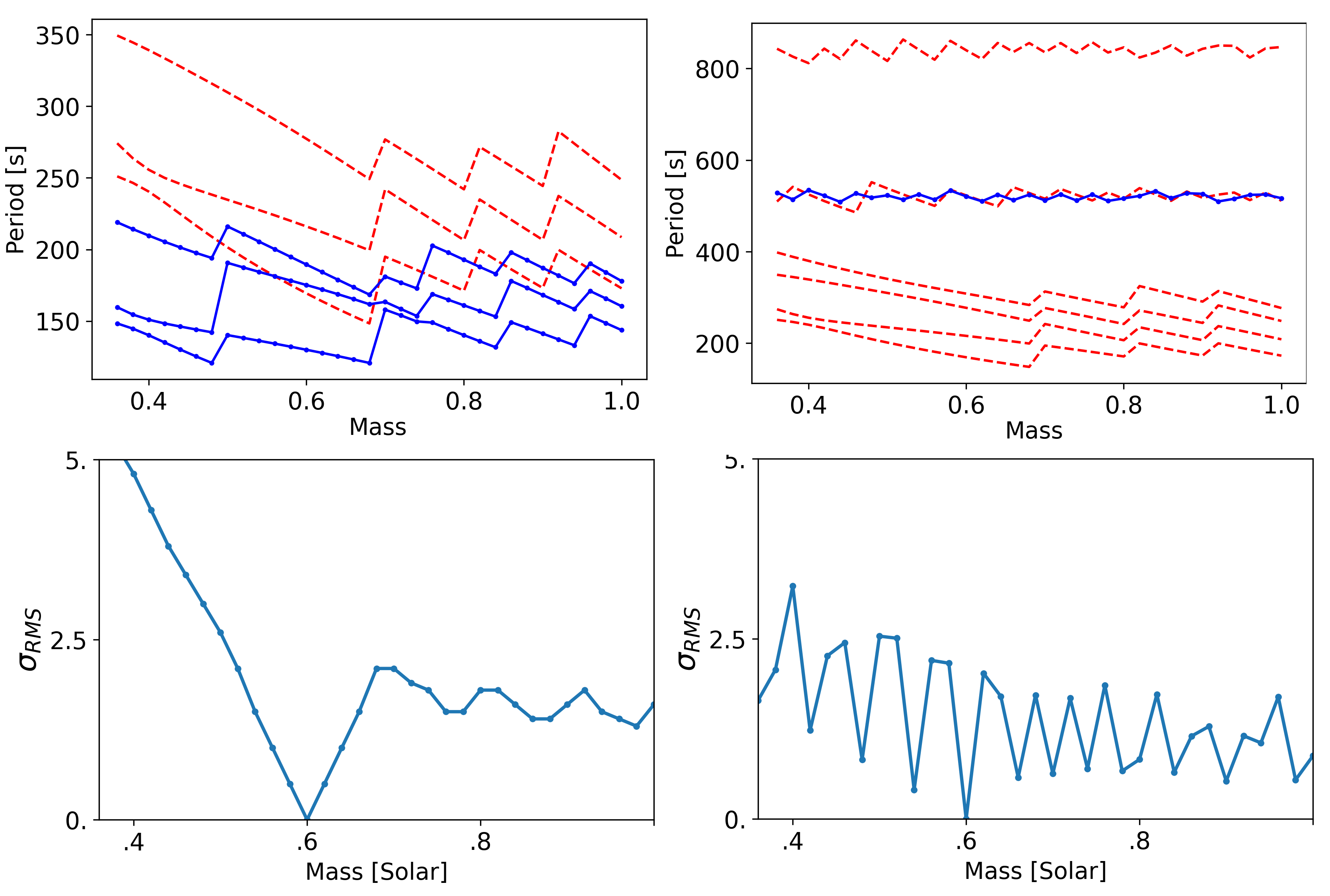}
\caption{Evolution of the best fit period for each mode of the GD 165-like period spectrum (left panels) and that of GD 66 (right panels) with changing mass. For each period spectrum, the corresponding dependence of the quality of fit ${\rm{\sigma_{RMS}}}$ is shown. \textbf{The solid blue lines correspond to $\ell=1$ modes, while the dashed red lines correspond ot $\ell=2$ modes.}
\label{fig:periodrifts_mass}
}
\end{figure}

\begin{figure}[h!]
\epsscale{0.7}
\plotone{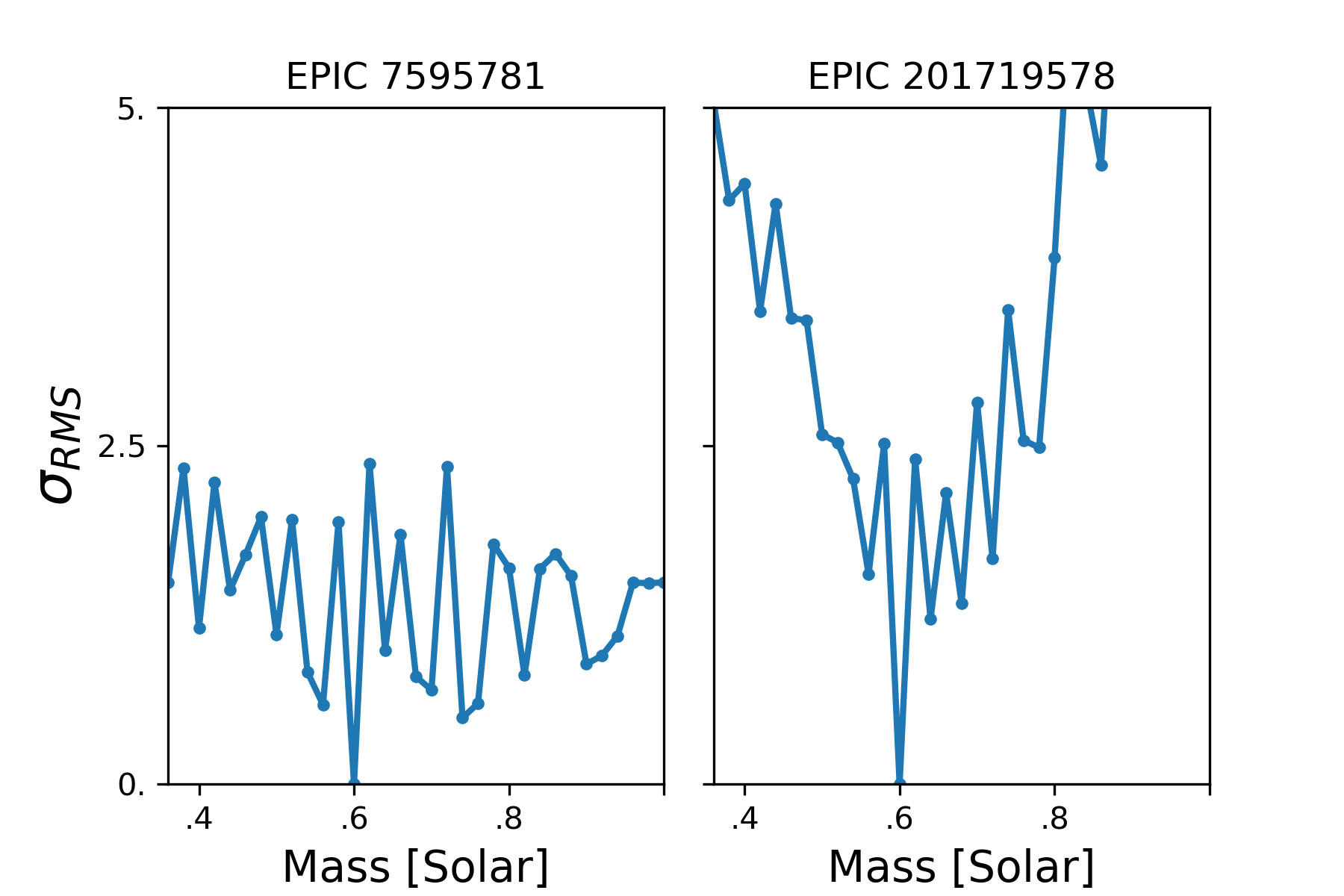}
\caption{Dependence of the quality of fit ${\rm{\sigma_{RMS}}}$ on mass for two rich pulsators. The pulsation spectrum based on that of EPIC 201719578 (right panel) has consecutive $\ell=1$ and $\ell=2$ mode sequences, while the other model only has a (shorter) $\ell=1$ sequence.
\label{fig:thetwotype1s_mass}
}
\end{figure}

\begin{itemize}
    \item The presence of longer period modes muddies the temperature determination. The reason for that is apparent in Fig. \ref{fig:periodrifts}, drawn for the richest pulsator considered in the study (EPIC 201719578, 23 periods). The period fitting algorithm always looks for the period matching the most closely. As the periods drift with the changing temperature, there are jumps to the adjacent k mode that occur as a result. Each one of these jumps lead to a reset of the goodness of fit ${\rm \sigma_{RMS}}$ to a lower value. The drift for the higher periods is steeper leading to multiple resets over the range of effective temperature considered. This produces the sawtooth patterns seen in Fig. \ref{fig:pardep}. When only shorter period modes are present ($\lesssim 700~s$), few if any resets occur.
    \item Similarly, the presence of long period modes muddies the mass determination. Fig. \ref{fig:periodrifts_mass} contrasts the period spectrum of a type 3 star (GD 165) with that of a type 4 object (GD 66). The only significant difference between the two period spectra is the presence of higher period modes in GD 66. The drift in periods is steeper with changing mass, requiring more frequent resets. For GD 165, there is a clean absolute minimum in mass. For GD 66, the best fitting mass is less well determined. Type 1 objects EPIC 7595781 and EPIC 201719578 have in common that they exhibit a wealth of long periods in their spectra, and yet, while EPIC 7595781 has a poorly determined mass, the mass for EPIC 201719578 suffers from less ambiguity (Fig \ref{fig:thetwotype1s_mass}). This could be due to the fact that the latter has such a rich period spectrum that the asymptotic period spacing \citep{Unno89} helps the mass determination. The fiducial model based on EPIC 201719578's period spectrum has 10 consecutive $\ell=1$ modes and 8 consecutive $\ell=2$ modes. EPIC 7595781's like spectrum has only 6 consecutive $\ell=1$ modes and lacks a continuous $\ell=2$ sequence of modes.
\end{itemize}

\begin{figure}[h!]
\epsscale{1.}
\plotone{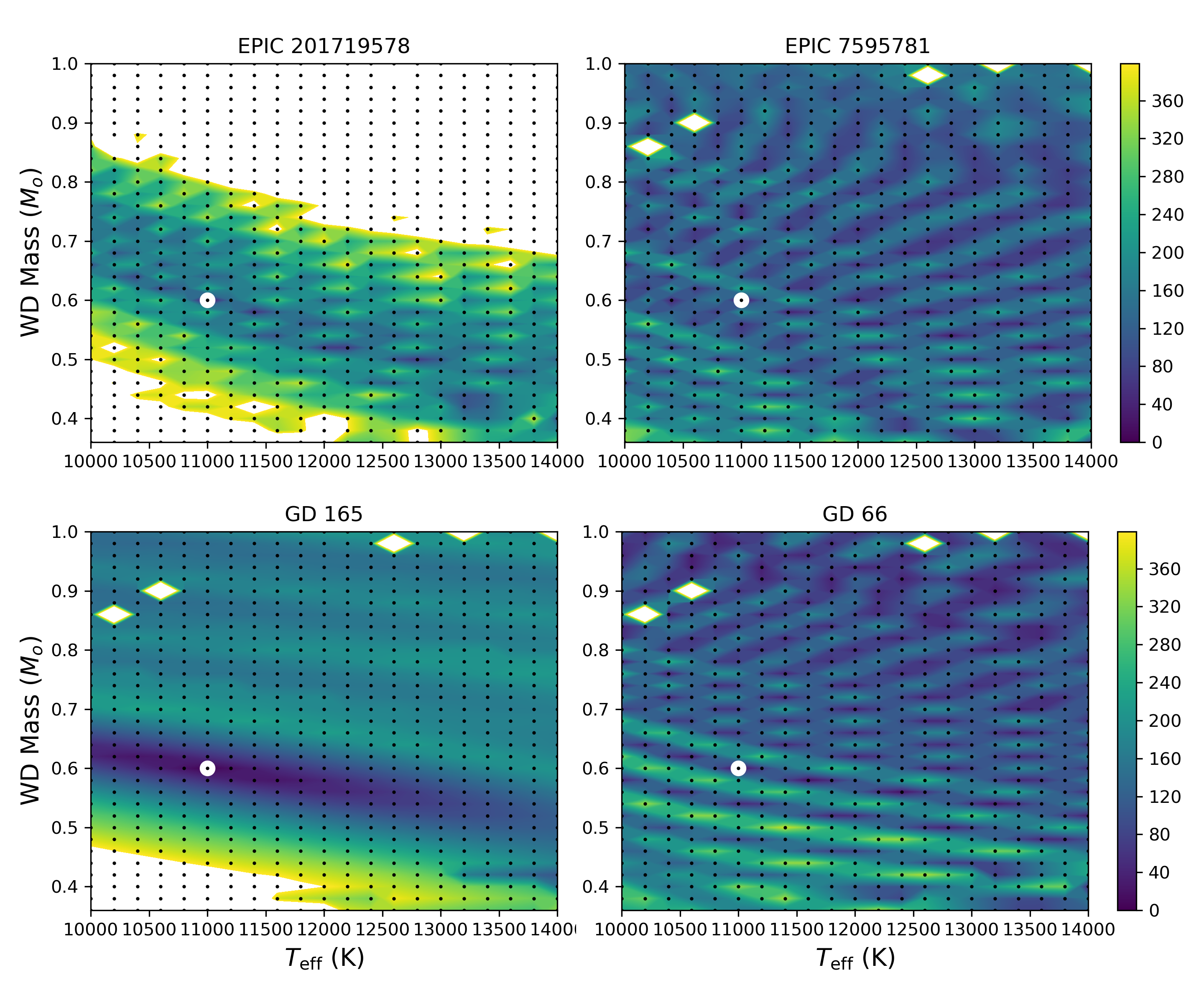}
\caption{Contour maps showing the quality of fit in the mass-effective temperature plane for four chosen periods spectrum extracted from the fiducial model. Each plot is labeled with the star that is mimicked in the selection of periods we fit from the fiducial model. The color scale is in 10ths of seconds and codes the fitness parameter (Eq. \ref{eq:fiteq1}). Zero is a perfect match. The dots are grid points. The white diamonds are gaps in the grid (5 models total). White regions have fitness parameters that go off the scale (very poor fits). The white circle marks the location of the best fit. All parameters, other than mass and effective temperature were held fixed, to the values listed in table \ref{tab:parameters}.  
\label{fig:contourplots_fid}
}
\end{figure}

We show the aggregate effect of what we discuss above in Fig. \ref{fig:contourplots_fid}. The figure shows contour maps of best fits in the mass-effective temperature plane, of the sort commonly found in the literature. We stress that these plots were produced by matching lists of periods that came from \emph{a fiducial model}. The model is on the grid, resulting in a unique, perfect fit and regular patterns. Another thing to emphasize is that the grid used in producing these plots only had varying effective temperature and mass. All other parameters were fixed to the values shown in table \ref{tab:parameters}. The two panels on the right show poorly determined mass and effective temperature. The GD 66 and EPIC 7595781-like period spectra have in common long period modes. The EPIC 201719578- and GD 165-like period spectra lead to better constrained mass and effective temperature for different reasons. GD 165 lacks long period modes, while EPIC 201719578 has long, consecutive sequences of $\ell=1$ and $\ell=2$ modes.

To get a sense of how that might apply in the asteroseismic fitting of real objects, we end this subsection by considering an actual fit, though one where we limit ourselves to varying two parameters, mass and effective temperature. We use a master grid of DAV models to fit GD 66's periods. The grid is coarse but sufficient in resolution and it covers much of the relevant parameters space for DAVs. The goal is to find a model that matches GD 66's observed period reasonably well, and fix all parameters save for mass and effective temperature to these values. This allows us to extract the effect of mass and effective temperature alone, in the fitting of an observed period spectrum. We perform the exercise twice: 1) using all of the periods observed, and 2) limiting ourselves to the four $\ell=1$ modes below 500~s. When choosing the best fit model on the master grid, we select the best fitting model within a mass and temperature range consistent with the spectroscopy on GD 66. This is reflective of what is typicaly done in forward asteroseismic fitting.

\begin{figure}[h!]
\epsscale{1.}
\plotone{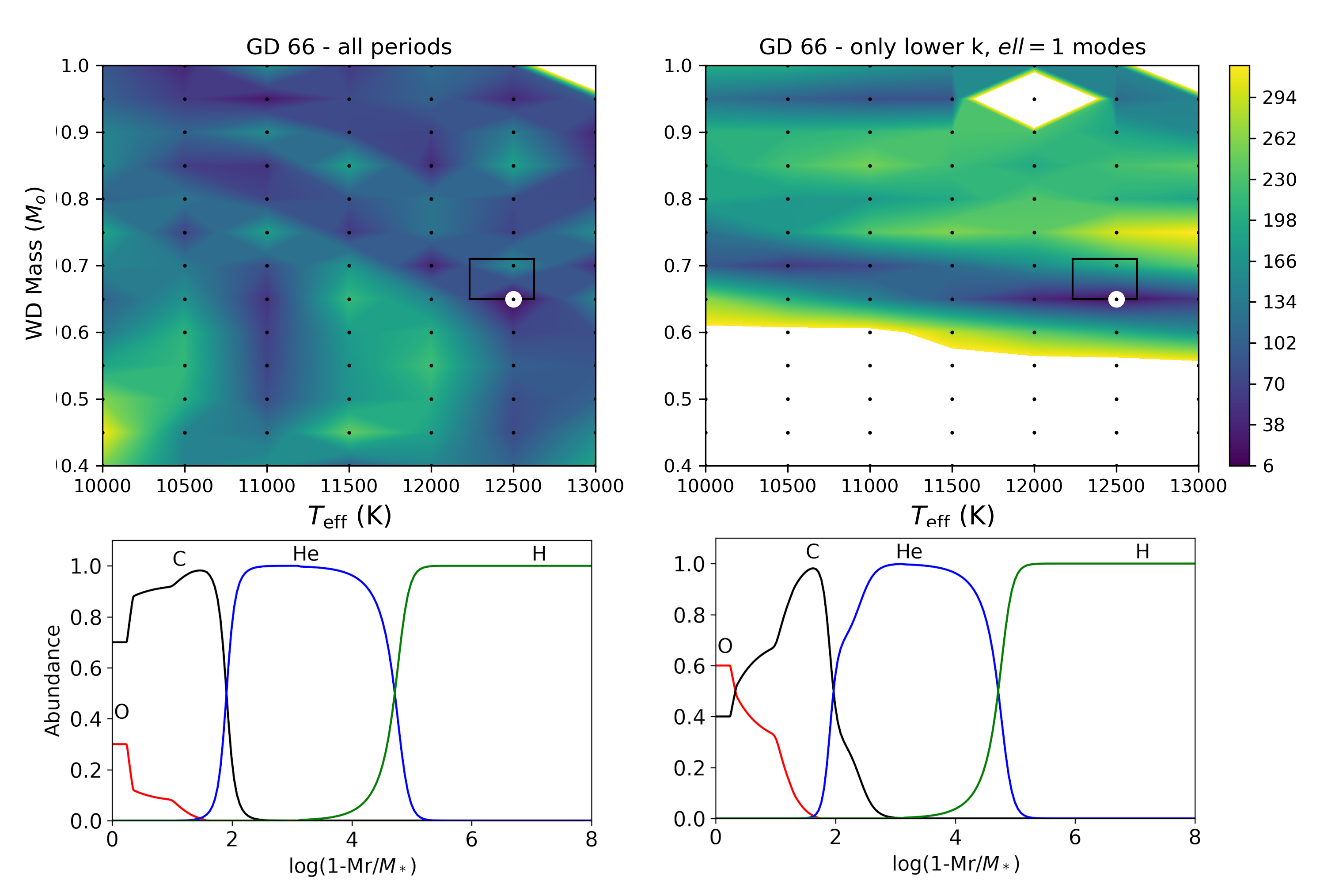}
\caption{Contour maps showing the quality of fit in the mass-effective temperature plane for GD 66's observed period . As in Fig. \ref{fig:contourplots_fid}, the color scale is in 10ths of seconds and codes the fitness parameter. The white diamond is a gap in the grid. White regions have fitness parameters that go off the scale (very poor fits). The white circle marks the location of the best fit. All parameters, other than mass and effective temperature were held fixed, but are different for each. The corresponding chemical profiles are shown below each contour map.
\label{fig:gd66}
}
\end{figure}

The result of this exercise is shown in Fig. \ref{fig:gd66}. Along with the best fit contour maps, we also graph the internal chemical profiles of the best fit models in each case. In both cases, the best fit model has the same mass and effective temperature because we constrained them to be so (there are only two grid points within the spectroscopic box). But the inclusion of the higher period modes affects what internal structure parameters best fit the period spectrum for that chosen mass and effective temperature.

\subsection{Envelope parameters}
\label{sec:discussion_envelope}
\begin{figure}[h!]
\epsscale{0.6}
\plotone{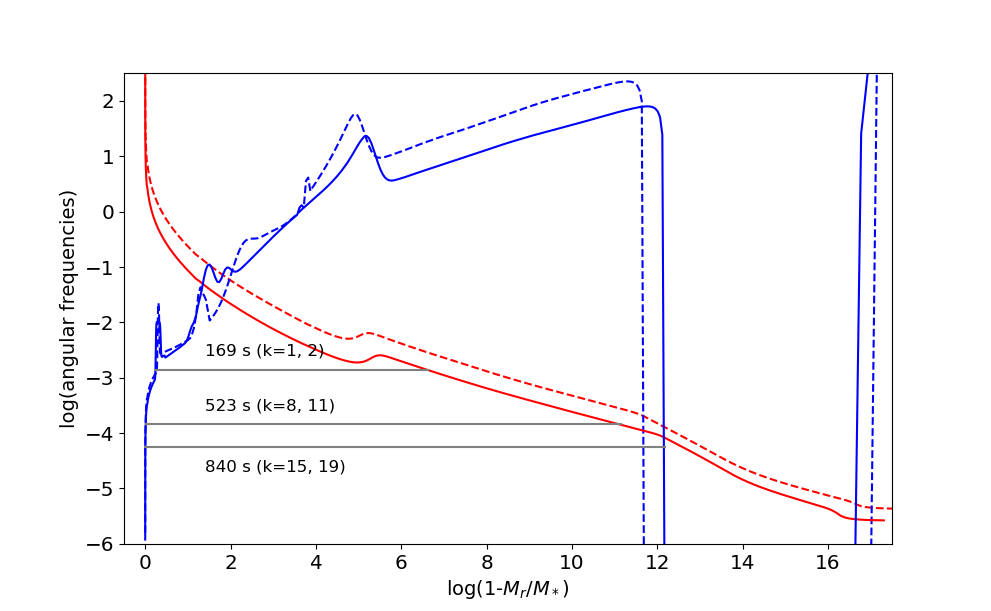}
\caption{The propagation diagram for the fiducial model (solid curves, Table \ref{tab:parameters}), with three key modes for the same indicated. The dashed lines are the \bvf frequency and Lamb frequency for a more massive, 0.8 \msun model. The "gap" in the \bvf curves between $\sim$ 12 and 16.5 corresponds to a region of the model that is convective. For each mode shown, we include in parenthesis its radial overtone for the fiducial model and that of a corresponding period in the 0.8 \msun model's period spectrum, respectively.
\label{fig:prop_diagram}
}
\end{figure}

\begin{itemize}
    \item Several of the envelope parameters are sensitive to all of the period spectra considered in this work and should be included in asteroseismic fitting as a matter of course. They include the helium envelope parameters ${\rm M_{env}}$ and ${\rm M_{He}}$, and to a lesser extent xhebar. On the flip side, the parameter that sets the sharpness of the transition at the base of the helium layer (alph1) is not as important. And the parameter that sets the sharpness of the transition to pure helium can safely be fixed (16 is a good value).   
    \item Much of the discussion about mass also applies to the thickness of the hydrogen layer ${\rm M_H}$. Long periods lead to multiple minima, consecutive sequences of modes help constrain that parameter. EPIC 211914185 is particularly sensitive to ${\rm M_H}$ because it only has 2 modes, and one of them (the $k=1$ mode) is trapped in the hydrogen layer (Fig. \ref{fig:weight_functions}).
    \item For most period spectra, it is not worth pushing ${\rm M_H}$ to values smaller than $\sim 10^{-7}$. 
    \item Convective efficiency only matters when fitting modes greater than $\sim 800$ s. It is not surprising that period spectra lacking any periods above 800~s are completely insensitive to convective efficiency. The resonant cavity of lower $k$ modes resides far from the base of the convection zone for any reasonable value of alpha (Fig. \ref{fig:prop_diagram}).
\end{itemize}

\subsection{Core parameters}
\label{sec:discussion_core}
\begin{figure}[h!]
\epsscale{0.8}
\plotone{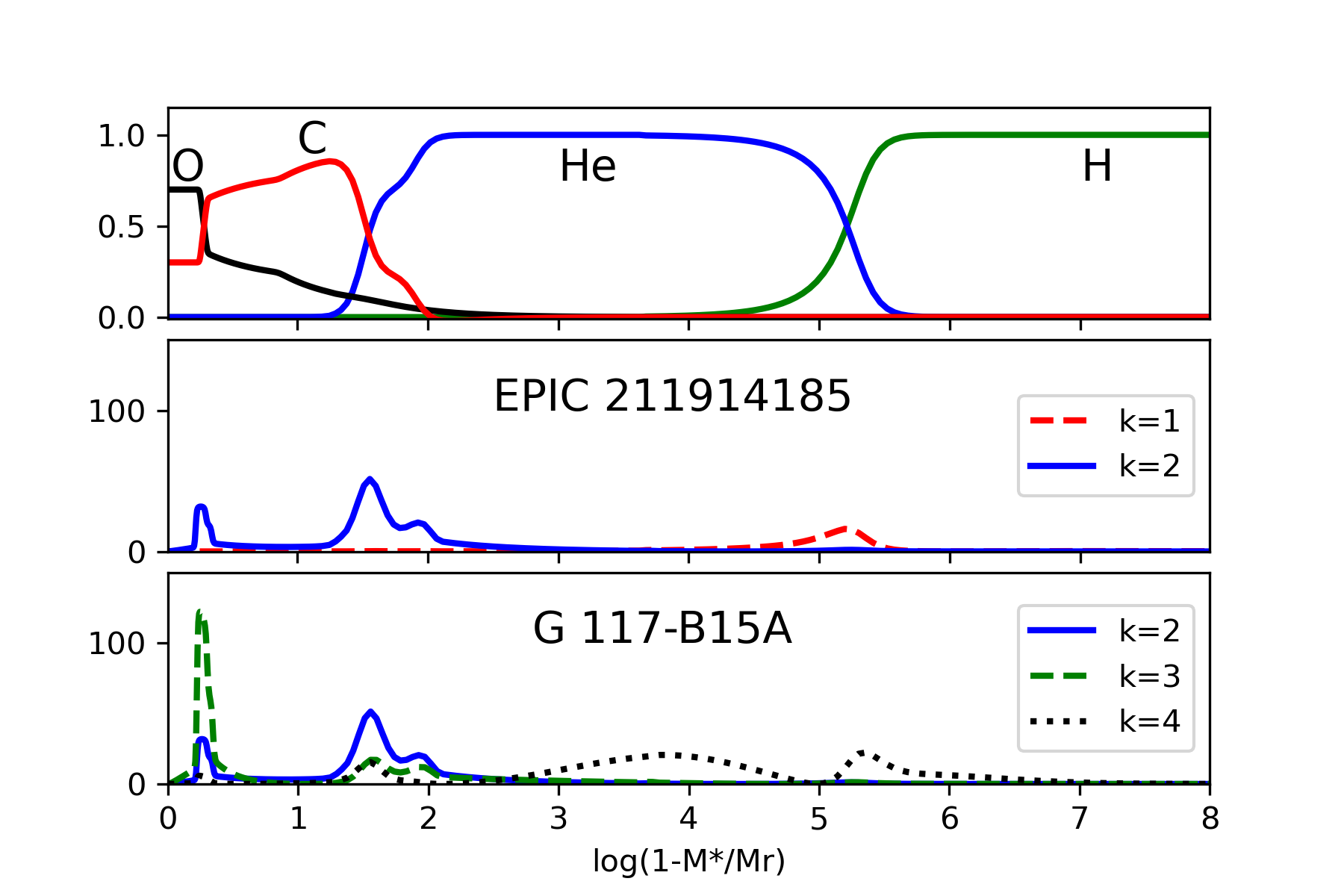}
\caption{Weight functions for the EPIC 211914185-like pulsation spectrum and the same for G 117-B15A. The former is observed to be insensitive to the shape of the oxygen abundance profile, while the latter is highly sensitive to it.
\label{fig:weight_functions}
}
\end{figure}

\begin{itemize}
    \item The core parameters h3, w2, and w3 universally matter little and can be fixed. The other three core parameters (h1, h2, and w1) matter at the level of ${\rm M_{env}}$ or Xhebar for the vast majority of period spectra. The only period spectrum mostly insensitive to any core structure is one where we include only two $\ell=1$ modes in the fitting ($k=1$ and $k=2$).   
    \item While the 2 mode spectrum described above is mostly insensitive to core structure, the G 117-B15A- like spectrum, also containing few $\ell=1$, low $k$ modes, responds strongly to core structure. It is among the most sensitive. Not all low-$k$ modes are created equal. In Fig. \ref{fig:weight_functions} we show the weight functions for the EPIC 211914185-like pulsation spectrum and the same for G 117-B15A. Weight functions provide a measure of how strongly the mode is affected by a given transition zone in the chemical profiles \citep{Montgomery03}. We identify the $k=3$ mode as responsible for the difference in sensitivity to the shape of the oxygen abundance profile. It is present in G 117-B15A-like fit, but not in the other.
    
\end{itemize}

\subsection{Mass dependence}
\label{disc_mass_dep}

We chose as our fiducial model an average mass of 0.6 ${\rm M_\odot}$. But as can be seen in table \ref{tab:objects}, masses of white dwarfs can vary. To understand how the results above transfer to lower or higher mass white dwarfs, it is worth examining the propagation diagram (\ref{fig:prop_diagram}). What determines the sensitivity of the modes to the internal structure of the model is the size of the their resonant cavity. In that respect, we see that the 169\:s mode  has an outer turning point a little further out in the 0.8 \msun model. But we also see that this does not make a difference in terms of which features of the \bvf frequency are inlcuded in the resonant cavity of the mode. The mode travels through the C/O core, and out past the helium/hydrogen transition zone. We do not expect that mode to have differing levels of sensitivity to the parameters that set the shape of those regions. A more important difference can exist for the longer period modes, which can have their outer turning point at the base of the convection zone. The depth of the convection zone comes into play, even though it does not make a very big difference. The $\sim$ 525 s modes stirs clear of the convection zone for each model, while the 840 s mode meets the base of the convection zone for both models. For a more thorough discussion of the calculation of the depth of the convection zone in WDEC and its dependence on mass, see \citet{Bischoff-Kim18a}. One key point from that discussion is that the depth of the convection zone is similar for different masses, both for DAVs and DBVs. While a mass dependent observational mapping of this is lacking, the dichotomy observed in mode stability in \citet{Hermes17} does appear to have a sharp edge at $\sim$ 800 s, with a sample of pulsating white dwarfs of varying masses. The implication of the above is that the results we obtain for a 0.6 \msun fiducial model can be applied to white dwarfs of differing masses. We also note that aside for the very low radial overtones (where the presence or absence of a given mode can make or break core sensitivity), the sensitivity of the modes to the internal structure of the models is a matter of the period, not of it's $k$ identification.

\subsection{ell identifications}
\label{disc_ell_id}

Just as the period itself matters more than its $k$ identification (with the caveat cited above), it also matters more than its $\ell$ identification. The exception to this statement is the presence of a significant number of $\ell=2$ modes in addition to $\ell=1$ modes. If the periods are in the asymptotic limit of even period spacing \citep{Unno89}, then having two sequences can help pin down the mass and effective temperature of the star.

\section{Conclusions}
\label{sec:conclusions}

In light of what was discussed, we can simplify the picture presented in table \ref{tab:objects} and focus on stars that present 1) short period modes $\lesssim$ 500 s, 2) long period modes $\lesssim$ 800 s and "surface modes", for the longer periods. In that scheme, types 2 and 3 merge, and so do types 6a and 6b. The parameter dependence plots of EPIC 210397465 (type 6a) and EPIC 229228364 (type 6b) are indeed very similar. When short period modes are present, sensitivity to core structure can vary widely. It is safer in any pipeline fitting to include the core parameters mentioned in section \ref{sec:discussion_core}. Some pulsation spectra with short period modes will strongly respond to core structure and some will not.

Perhaps the most important and unexpected result that comes out of this work is that more periods does not always mean better constrained fits. Most saliently, while longer period modes are more sensitive to the effective temperature and mass of the model, this very sensitivity also leads to multiple local minima. On the flip side, any star with a long sequence of either $\ell=1$ or $\ell=2$ modes, or better yet both, should yield a well determined, unique solution in the mass-effective temperature plane. Another important result is the lack of a perfect correlation between the presence of low-$k$ modes and sensitivity to core parameters. When there is a very small number of low-$k$ modes present, sensitivity can vary widely. This is not a new finding, but it is worth repeating.

It may be tempting to cherry pick periods (for instance ignore longer period modes, or only include known multiplets for $\ell$ identification) when performing an asteroseismic fit but we strongly caution against that, as it can lead to precise, but wrong parameter determinations. For GD 66, we had to pick different core structure parameters in order to constrain the best fits to land within the spectroscopic box. This is an important point, as it is common (and fair), in forward asteroseismic fitting, to give more weight to the best fits that land within spectroscopic boxes. In this work we performed a numerical experiment, partly to show what not to do. Every known, independent mode that clearly rises above the noise should be included in any asteroseismic fitting.

Aside for the effect of longer period modes on how well determined some parameters are, the number of modes does not come into play in the types of numerical experiments we did here, as we varied only one parameter at a time. When multiple parameters are varied at once, one has to take into consideration the interplay between the parameters and watch out for under-constrained fits.

This work can inform future efforts at pipeline fitting of pulsating white dwarfs, especially with the WDEC, as it is the specific parameterization of that code that we used. However, we also presented more general principles that are more widely applicable. 

\software{WDEC \citep{Bischoff-Kim18b}, https://github.com/kim554/wdec)}

\begin{appendix}
For reference, we provide below the complete list of periods for the fiducial model used in this work, along with the observed periods of each star listed in table \ref{tab:objects}. We lined up the periods with periods in the fiducial model. Periods that have an asterisk next to them are members of multiplets. This helps determine their $\ell$ identification. This is by no means a formal fit of any of the objects, nor a formal determination of mode identifications. The object is to inform a choice of period spectra representative of what one might encounter in fitting pulsating DA white dwarf spectra. 

\begin{table}
  \begin{center}
  \caption{Complete list of periods for the fiducial model and observed period spectra (in seconds) for the stars considered in this work - types 1 and 4, $\ell=1$ modes.
     \label{tab:fid_periods_t1_t4_ell1}
}
  \begin{tabular}{|cc|cc|cc|}
  \hline
     &   & \multicolumn{2}{c}{Type 1}  & \multicolumn{2}{|c|}{Type 4}  \\
    k  & Fid. model& EPIC 7595781 & EPIC 201719578 & GD 66 & G 185-32 \\
    \hline
    \hline
    1  & 169.354  &         &           & 197.23*   & 141*  \\
    2  & 216.094  & 206.814 &           & 256*      & 216*  \\
    3  & 277.320  & 280.882*&           & 271.7144* &       \\
    4  & 308.317  & 295.981*&           & 302.7653* & 300*  \\
    5  & 354.882  & 356.86  & 368.624*  &           & 370*  \\
    \hline
    6  & 418.089  & 396.146*& 404.603*  &           &       \\
    7  & 476.030  & 480.335 & 461.088*  &           &       \\
    8  & 522.701  &         & 505.447   &  523.2533 &       \\
    9  & 577.442  &         & 557.244   &           &       \\
    10 & 614.553  &         &           &           &       \\
    \hline
    11 & 652.808  &         &           &           & 652   \\
    12 & 694.763  & 683.936 & 679.49    &           &       \\
    13 & 725.318  &         & 737.76    &           &       \\
    14 & 777.528  &         & 773.12    &           &       \\
    15 & 839.636  &         & 846.907   &  810.4    &       \\
    \hline
    16 & 894.814  &         & 903.1     &           &       \\
    17 & 939.483  &         & 922.6     &           &       \\
    18 & 983.983  &         & 966.34    &           &       \\
    19 & 1038.534 &         & 1014.6    &           &       \\
    20 & 1076.953 &         & 1051.3    &           &       \\
    \hline
    21 & 1118.446 & 1129.19 & 1095.434  &           &       \\
    22 & 1174.906 &         &           &           &       \\
    23 & 1213.114 &         &           &           &       \\
    24 & 1265.217 &         &           &           &       \\
    25 & 1317.029 &         &           &           &       \\
    \hline
    26 & 1364.226 &         &           &           &       \\
    27 & 1407.147 &         &           &           &       \\
    28 & 1471.843 &         &           &           &       \\
  \hline  
 \end{tabular}
 \end{center}
\vspace{1mm}
\end{table}

\begin{table}
  \caption{Complete list of periods for the fiducial model and observed period spectra (in seconds) for the stars considered in this work - types 1 and 4, $\ell=2$ modes.
     \label{tab:fid_periods_t1_t4_ell2}
}
\noindent
  \begin{tabular}{@{} |cc|cc|cc|cc|c}
  \hline
     &   & \multicolumn{2}{c}{Type 1}  & \multicolumn{2}{|c|}{Type 4}  \\
    k& Model& EPIC 7595781 & EPIC 201719578 & GD 66 & G 185-32 \\
    \hline
    \hline
    1  & 97.804   &         &           &           &                 \\
    2  & 129.934  &         &           &           & 148              \\
    3  & 175.180  &         &           &           &                  \\
    4  & 189.483  &         &           &           & 182             \\
    5  & 205.438  &         &           &           & 213              \\
    \hline
    6  & 242.230  & 261.213 &           &           &                 \\
    7  & 275.645  & 279.646 &           &           & 265              \\
    8  & 303.273  &         &           &           &                 \\
    9  & 338.069  & 328.5*  &           &           &                 \\
    10 & 363.538  & 350.322 &           &           &                 \\
    \hline
    11 & 387.466  &         &           &           &                  \\
    12 & 414.695  &         &           &           &                \\
    13 & 447.787  &         &           &           &                \\
    14 & 478.087  &         &           &           &               \\
    15 & 491.066  &         &           &           &                 \\
    \hline
    16 & 520.694  &         &           & 518.6029  &                \\
    17 & 548.586  &         &           &           &                \\
    18 & 574.515  &         &           &           &                \\
    19 & 604.214  &         &           &           &        \\
    20 & 639.280  &         &           &           &             \\
    \hline
    21 & 663.474  &         &           &           &                \\
    22 & 690.805  &         &           &           &                 \\
    23 & 721.172  &         &           &           &                  \\
    24 & 746.303  &         & 748.9     &           &                \\
    25 & 771.533  &         & 787.94    &           &               \\
    \hline
    26 & 802.906  &         & 799.896   &           &                \\
    27 & 824.510  &         & 820.2     &           &              \\
    28 & 853.262  &         & 834.99    &           &                \\
    29 & 887.816  &         & 861.4     &           &                 \\
    30 & 913.832  &         & 877.44    &           &                 \\
    \hline
    31 & 946.058  &         & 947.1     &           &                  \\
    32 & 975.743  &         &           &           &                \\
    33 & 1009.875 &         &           &           &                \\
    34 & 1031.534 &         &           &           &               \\
    35 & 1056.938 &         &           &           &                \\
  \hline  
 \end{tabular}

\vspace{1mm}
\end{table}

\begin{table}
  \begin{center}
  \caption{List of periods for the fiducial model and observed period spectra (in seconds) for the stars considered in this work - types 2 and 3.
     \label{tab:fid_periods_ell1_t2_t3}
}
  \begin{tabular}{|ccc|cc|cccc|}
  \hline 
     & &    & \multicolumn{2}{c|}{Type 2}  & \multicolumn{4}{c|}{Type 3}  \\
    $\ell$  & k  & Fid. model& EPIC 211914185 & G 117-B15A & EPIC 220258806 & R548 & L19-2 & GD 165\\
    \hline
    \hline
    1 & 1   & 169.354   & 109.15103*  &       & 119.87751   & 187   & 118.5191  & 120.36*   \\
    1 & 2   & 216.094   & 190.4543*   & 215.2 & 196.8408*   & 213*  & 143.4191  & 192.68*   \\
    1 & 3   & 277.320   &             & 270.5 & 251.6535    & 274*  & 192.6096* & 250.16*   \\
    1 & 4   & 308.317   &             & 304.1 & 298.2676*   &       &           &           \\
    1 & 5   & 354.882   &             &       & 349.913     & 334   & 350.15    &           \\
    \hline
    \hline
    2 & 1   & 97.804    &             &       & 116.3651*   &       &           &           \\
    2 & 2   & 129.934   &             &       & 146.336*    &       & 113.7779  & 114.23*   \\
    2 & 3   & 175.180   &             &       & 169*        &       &           & 146.32*   \\
    2 & 4   & 189.483   &             &       &             &       &           & 168.19    \\
    2 & 5   & 205.438   &             &       &             & 218   &           &           \\
    \hline
    2 & 6   & 242.230   &             &       &             &       &           &           \\
    2 & 7   & 275.645   &             &       & 270.8699    &       &           &           \\
    2 & 8   & 303.273   &             &       & 314.5611    & 318*  &           &           \\
    2 & 9   & 338.069   &             &       &             &       &           &           \\
    2 & 10  & 363.538   &             &       & 356.1452    &       &           &           \\
    \hline
 \end{tabular}
 \end{center}
\vspace{1mm}
\end{table}

\begin{table}
  \begin{center}
  \caption{List of periods for the fiducial model and observed period spectra (in seconds) for the stars considered in this work - types 5, 6a and 6b, $\ell=1$ modes.
     \label{tab:fid_periods_t5_t6_ell1}
}
  \begin{tabular}{|cc|cc|cc|}
  \hline 
     &   & \multicolumn{2}{c|}{Type 5} & Type 6a & Type 6b   \\
    k  & Fid. model  & EPIC 220204626  & EPIC 206212611   &     EPIC 210397465 & EPIC 229228364 \\
    \hline
    \hline
    6  & 418.089  &             &           &           &           \\
    7  & 476.030  &             &           &           &           \\
    8  & 522.701  & 508.0354*   &           &           &           \\
    9  & 577.442  & 582*        &           &           &           \\
    10 & 614.553  & 627.447     &           & 605.86    &           \\
    \hline
    11 & 652.808  & 676.385     &           & 667.0133* &           \\
    12 & 694.763  &             &           & 710.5527* &           \\
    13 & 725.318  &             &           &           &           \\
    14 & 777.528  & 794.8108*   &           & 758.4643  &           \\
    15 & 839.636  &             &           &           &           \\
    \hline
    16 & 894.814  &             &           &           &           \\
    17 & 939.483  &             &           &           &           \\
    18 & 983.983  &             &           & 977.055   &           \\
    19 & 1038.534 &             & 1033*     & 1019.00   &           \\
    20 & 1076.953 &             &           &           & 1073      \\
    \hline
    21 & 1118.446 &             &           &           & 1118.07   \\
    22 & 1174.906 &             &           &           &           \\
    23 & 1213.114 &             &           &           & 1204.841  \\
    24 & 1265.217 &             & 1237      & 1278.10   &           \\
    25 & 1317.029 &             &           &           &           \\
    \hline
    26 & 1364.226 &             &           & 1382.26   &           \\
    27 & 1407.147 &             &           &           &           \\
    28 & 1471.843 &             &           &           &           \\
  \hline  
 \end{tabular}
 \end{center}
\vspace{1mm}
\end{table}

\begin{table}
  \caption{Complete list of periods for the fiducial model and observed period spectra (in seconds) for the stars considered in this work - types 5 and 6a, $\ell=2$ modes.
     \label{tab:fid_periods_t5_t6a_ell2}
}
\noindent
  \begin{tabular}{|cc|cc|c|}
  \hline
     &   & \multicolumn{2}{c|}{Type 5}  & Type 6a  \\
    k& Model& EPIC 220204626 & EPIC 206212611 & EPIC 210397465   \\
    \hline
    \hline
    16 & 520.694  & 511.202*    &           &           \\
    17 & 548.586  &             &           &           \\
    18 & 574.515  &             &           &           \\
    19 & 604.214  &             &           &  600.53*  \\
    20 & 639.280  &             &           &           \\
    \hline
    21 & 663.474  &             &           &           \\
    22 & 690.805  &             &           &           \\
    23 & 721.172  &             &           &           \\
    24 & 746.303  &             &           &           \\
    25 & 771.533  &             &           &           \\
    \hline
    26 & 802.906  & 791.4630*    &           &           \\
    27 & 824.510  &             &           &           \\
    28 & 853.262  &             &           &           \\
    29 & 887.816  &             &           &           \\
    30 & 913.832  &             &           &           \\
    \hline
    31 & 946.058  &             &           &           \\
    32 & 975.743  &             &           &           \\
    33 & 1009.875 &             &           &           \\
    34 & 1031.534 &             &           &           \\
    35 & 1056.938 &             &           &           \\
    \hline
    36 & 1079.353 &             &           & 1072.39*  \\
    37 & 1114.519 &             &           &           \\
    38 & 1146.450 &             &           &           \\
    39 & 1177.655 &             &           &           \\
    40 & 1210.473 &             &           &           \\
    \hline
    41 & 1237.904 &             &           & 1220.306  \\
    42 & 1267.144 &             &           &           \\
    43 & 1288.403 &             &           &           \\
    44 & 1319.872 &             &           &           \\
    45 & 1344.115 &             & 1336.39*  &           \\
    \hline
    46 & 1378.940 &             &           &           \\
    47 & 1406.465 &             &           &           \\
    48 & 1438.517 &             &           &           \\
    49 & 1463.892 &             &           &           \\
    50 & 1496.537 &             &           &           \\
  \hline  
 \end{tabular}

\vspace{1mm}
\end{table}

\end{appendix}
\bibliography{index}{}
\bibliographystyle{aasjournal}

\end{document}